\documentclass[10pt,conference]{IEEEtran}
\IEEEoverridecommandlockouts

\usepackage[dvipsnames,table,xcdraw]{xcolor}

\usepackage{url} 
\usepackage{framed}
\usepackage{mdframed}
\usepackage{listings}
\usepackage{multicol}
\usepackage{amssymb}
\usepackage{lipsum}
\usepackage{adjustbox}
\usepackage[most]{tcolorbox}
\usepackage[font=bf,skip=\baselineskip]{caption}
\usepackage{tabularx}
\usepackage{seqsplit}
\usepackage{svg}
\usepackage{float}
\usepackage{multirow}
\usepackage{booktabs}
\usepackage{graphicx}
\usepackage{subcaption}
\usepackage{makecell}
\usepackage[ruled,vlined,linesnumbered]{algorithm2e}
\usepackage{amsmath}
\usepackage{tikz}
\usepackage{threeparttable}
\usepackage{breakurl} 
\usepackage{boldline}
\usepackage{pifont}
\usepackage{balance}

\usepackage{enumitem}
\usepackage[htt]{hyphenat}
\usepackage{array}

\definecolor{customblue}{HTML}{006ca6}
\definecolor{customgreen}{HTML}{009264}
\definecolor{custombrown}{HTML}{ff3d00}
\AtEndPreamble{
\usepackage{hyperref}

    \hypersetup{
      colorlinks = true,
      linkcolor = customgreen,
      anchorcolor = purple,
      citecolor = customgreen,
      filecolor = purple,
      urlcolor = customgreen
    }
}

\newcommand{\approach}{CKGFuzzer\xspace} 
\begin{document}

\title{\approach: LLM-Based Fuzz Driver Generation Enhanced By Code Knowledge Graph}

\author{
\IEEEauthorblockN{Hanxiang Xu\IEEEauthorrefmark{1}\IEEEauthorrefmark{5}, Wei Ma\IEEEauthorrefmark{2}\IEEEauthorrefmark{5}, Ting Zhou\IEEEauthorrefmark{1}, Yanjie Zhao\IEEEauthorrefmark{1}\IEEEauthorrefmark{6}, Kai Chen\IEEEauthorrefmark{1}, Qiang Hu\IEEEauthorrefmark{3}, Yang Liu\IEEEauthorrefmark{4} and Haoyu Wang\IEEEauthorrefmark{1}}

\IEEEauthorblockA{\IEEEauthorrefmark{1}
Huazhong University of Science and Technology, Wuhan, China\\
xuhx@hust.edu.cn, tingzhou27@hust.edu.cn, yanjie\_zhao@hust.edu.cn, kchen@hust.edu.cn, haoyuwang@hust.edu.cn}
\IEEEauthorblockA{\IEEEauthorrefmark{2}
Singapore Management University, Singapore, 
weima93@gmail.com}
\IEEEauthorblockA{\IEEEauthorrefmark{3}
The University of Tokyo, Tokyo, Japan, qianghu0515@gmail.com}
\IEEEauthorblockA{\IEEEauthorrefmark{4}
Nanyang Technological University, Singapore, 
yangliu@ntu.edu.sg}
\thanks{\IEEEauthorrefmark{5} Hanxiang Xu and Wei Ma contributed equally to this work.}
\thanks{\IEEEauthorrefmark{6} Yanjie Zhao is the corresponding author.}
}

\maketitle

\begin{abstract}
In recent years, the programming capabilities of large language models (LLMs) have garnered significant attention. Fuzz testing, a highly effective technique, plays a key role in  enhancing software reliability and detecting vulnerabilities. However, traditional fuzz testing tools rely on manually crafted fuzz drivers, which can limit both testing efficiency and effectiveness. To address this challenge, we propose an automated fuzz testing method driven by a code knowledge graph and powered by an LLM-based intelligent agent system, referred to as \approach. We approach fuzz driver creation as a code generation task, leveraging the knowledge graph of the code repository to automate the generation process within the fuzzing loop, while continuously refining both the fuzz driver and input seeds. The code knowledge graph is constructed through interprocedural program analysis, where each node in the graph represents a code entity, such as a function or a file. The knowledge graph-enhanced \approach 
not only effectively resolves compilation errors in fuzz drivers and generates input seeds tailored to specific API usage scenarios, but also analyzes fuzz driver crash reports, assisting developers in improving code quality. By querying the knowledge graph of the code repository and learning from API usage scenarios, we can better identify testing targets and understand the specific purpose of each fuzz driver. We evaluated our approach using eight open-source software projects. The experimental results indicate that \approach achieved an average improvement of 8.73\% in code coverage compared to state-of-the-art techniques. Additionally, \approach reduced the manual review workload in crash case analysis by 84.4\% and successfully detected 11 real bugs (including nine previously unreported bugs) across the tested libraries. Our research enhances the overall performance of fuzz testing by refining fuzz driver generation strategies and input seed analysis, offering a more effective solution for vulnerability remediation and software quality improvement.

\end{abstract}

\section{Introduction}

Since the emergence of Large language models (LLMs), their remarkable programming capabilities have attracted significant attention from researchers. 
By leveraging the code generation capabilities of LLMs, researchers are exploring solutions to more complex programming challenges~\cite{10.1145/3695988}, such as software self-repair and code translation. Among these, fuzzing has gained significant attention due to its prominent role in identifying software vulnerabilities and enhancing software reliability. Consequently, an increasing number of studies are utilizing LLMs to further improve the performance of fuzzing in order to better address these challenges.
Although traditional fuzzing tools, such as AFL~\cite{AFL} and libFuzzer~\cite{llvm_libfuzzer}, are well-established, the creation of fuzz drivers still relies heavily on manual coding, which limits the efficiency and scope of testing. As software systems grow in size and complexity, manually writing fuzz drivers becomes increasingly time-consuming and prone to overlooking potential test scenarios. Given that the effectiveness of fuzz testing is closely tied to the quality of the fuzz drivers, low-quality drivers can significantly weaken the effectiveness of the tests. Additionally, the input seeds used in fuzz testing play a crucial role in determining the success of the process.

Existing approaches, such as PromptFuzz~\cite{lyu2023prompt}, generate fuzz drivers using LLMs through different combinations of API mutations. These methods employ zero-shot or few-shot learning techniques, such as Google OSS-Fuzz-Gen\footnote{\url{https://github.com/google/oss-fuzz-gen}}, to generate fuzz drivers based on a given set of APIs. Zhang et al.~\cite{zhang2024survey} have explored various generation strategies as well. However, these methods have not fully addressed the challenges associated with generating high-quality fuzz drivers, resulting in limitations in both the quality and effectiveness of the generated drivers. For example, while APIs in libraries can be called independently, they often have specific use cases, and different functional requirements may necessitate combining various APIs. Recognizing these combinations is crucial for developers when analyzing crashes, as it helps trace the source of errors and assess their impact, ultimately improving the efficiency of vulnerability fixes and software quality. Moreover, fuzz driver generation differs from unit testing; it not only requires an understanding of individual functions but also the interactions between multiple APIs, demanding deep insight into system code behavior. Without sufficient contextual information, LLMs struggle to generate high-quality fuzz drivers.

Another key challenge lies in fuzzing's heavy reliance on input seeds. To produce more meaningful input seeds, data flow analysis on the fuzz driver is required. However, achieving this with static or dynamic analysis tools increases engineering complexity. Crash analysis presents yet another challenge, particularly in distinguishing between API misuse and library-level errors. Crashes may stem from the incorrect fuzz driver, improper API usage in the fuzz driver or deeper issues within the system, making it challenging for developers to accurately trace the root cause. This uncertainty complicates the debugging process and may hinder the effective resolution of vulnerabilities. Additionally, LLM-based crash analysis depends on comprehensive contextual information and up-to-date knowledge of vulnerabilities, which are not always readily available, further complicating the identification of potential issues.

To address these challenges, we approach fuzz driver generation as a code generation task. By constructing a knowledge graph based on code repositories, we use LLMs to automatically generate fuzz drivers and perform data flow analysis to produce input seeds, thereby achieving fully automated fuzz testing. To enable LLMs to better understand the software under test, we propose abstracting the codebase into a code knowledge graph. In this graph, nodes represent code elements (such as functions, classes, and modules), while edges depict their interactions, with each node containing detailed information about the element’s functionality and behavior. We treat fuzz driver generation as a code planning task, beginning with the formulation of a generation plan followed by iterative fuzz driver generation. 
To this end, we developed an LLM-driven fuzz driver agent system, \approach, based on the code knowledge graph. Our approach fully automates the fuzzing process, significantly reducing engineering complexity while improving the usability of fuzz testing. The code knowledge graph provides deeper insights into code behavior, guiding the LLM in generating higher-quality API combinations for fuzz drivers and coverage-guided mutation.

We evaluated our approach using eight open-source libraries and addressed three research questions~(RQs). First, we compared \approach with other generation techniques~(RQ1). Second, we analyzed the influence of each agent in \approach~(RQ2). Third, we investigated whether \approach can effectively analyze the causes of crashes generated during fuzzing~(RQ3). Our experiments demonstrate that \approach, supported by the code knowledge graph, can generate high-quality fuzz drivers with enhanced code coverage. Through extensive evaluations across these libraries, we found that the agents of \approach play a crucial role in fuzz driver generation. Additionally, our analysis of crash reports indicates that \approach can provide valuable insights for developers.

In summary, our work has three main contributions:

\begin{enumerate}
    \item We propose \approach, which leverages a code knowledge graph to enhance fuzz driver generation. By analyzing the relationships between code elements and APIs, \approach automatically generates high-quality fuzz drivers that achieve better code coverage.
    \item \approach introduces a fully automated fuzzing framework that integrates multi-agent systems and LLMs. It automates key steps such as fuzz driver generation, input seed initiation, dynamic program repair, and coverage-guided mutation, significantly reducing manual effort and enhancing overall efficiency in the fuzzing process.
    \item Experimental results show that \approach achieved an average increase of 8.73\% in code coverage compared to the state-of-the-art methods. Furthermore, \approach reduced the manual review workload in crash case analysis by 84.4\% and successfully identified 11 real bugs (including nine previously unreported bugs) of tested libraries. 
\end{enumerate}

\noindent \textbf{Artifact Availability.} We make our artifact public available at \url{https://github.com/security-pride/CKGFuzzer}.

\section{Background and Related Work}
\subsection{Fuzz Testing}
Fuzzing is a technique used to detect software security vulnerabilities~\cite{zhao2023understandingprogramsexploitingfuzzing}. It works by strategically mutating input data and executing test cases while monitoring the execution process, thereby effectively identifying defects. However, despite the high efficiency of fuzz testing in detecting flaws, it faces a significant challenge: the input space is vast, while the inputs that can trigger defects are exceedingly rare. As a result, the core of fuzzing lies in how to efficiently identify these sparse inputs that may cause software vulnerabilities. To increase the likelihood of discovering defects, fuzzers often use execution feedback (such as execution states or outcomes) as a fitness metric, with code coverage being a common example. Depending on the level of analysis performed on the system under test, fuzzing techniques can be categorized into three types~\cite{li2018fuzzing}: black-box, white-box, and gray-box. Black-box fuzzing depends solely on external observations without internal insights, while white-box fuzzing requires detailed analysis of the source code, with gray-box fuzzing positioned as an intermediate approach between the two.

Fuzzing has a wide range of applications~\cite{10.1145/3623375, manes2018art,10.1007/978-3-642-39611-3_28,10.1145/3243734.3243804}, including grammar-based test case generation,  kernel testing, GUI testing and penetration testing. One of the most well-known and versatile fuzzing tools is American Fuzzy Lop (AFL). Additionally, there are tools tailored to specific domains, such as Bbuzz~\cite{8170785}, which focuses on network protocol testing, and Ffuf~\cite{ffuf}, designed for web-based testing. Library API fuzzing is an important application area, focusing on testing the security of library interfaces. Compared to general fuzzing methods, Library API fuzzing targets library interfaces specifically, aiming to uncover potential vulnerabilities in how they handle exceptional inputs. This method is commonly used to test libraries that are widely used across different applications, ensuring their security and stability under various input conditions. LibFuzzer is a widely-used tool for testing library APIs, generating large volumes of random input to evaluate the robustness of library interfaces. Several other tools also integrate code analysis techniques to enhance fuzzing. For example, RULF~\cite{10.1109/ASE51524.2021.9678813} tests Rust libraries by analyzing API dependency graphs, ensuring safe operation in complex dependency contexts, while GraphFuzz~\cite{10.1145/3510003.3510228} leverages data flow analysis for API testing to uncover potential security risks.
\subsection{LLM for Fuzzing}
The integration of LLMs with fuzzing has received significant attention in recent research, particularly in the context of generating more intelligent and context-aware test cases for software systems~\cite{huang2024largelanguagemodelsbased,xu2024large,10.1145/3663529.3663784}. Traditional fuzzing techniques often struggle with generating meaningful inputs for complex software systems. LLMs, with their powerful generative capabilities, have shown great promise in addressing these limitations by enabling more sophisticated fuzzing strategies. In recent years, several LLM-based fuzzers have been proposed to improve the efficiency and effectiveness of fuzzing techniques. For example, LLMs have been used in tools like TitanFuzz~\cite{deng2023large}, FuzzGPT~\cite{deng2024large}, WhiteFox~\cite{yang2023whitefox} and ParaFuzz~\cite{yan2024parafuzz}, which leverage models such as GPT~\cite{achiam2023gpt}, Codex~\cite{finnie2022robots}, and other generative models to create diverse and high-quality test inputs. These fuzzers introduce LLMs into prompt engineering and seed mutation to enhance fuzzing performance.

LLMs are also used to automate the fuzzing process. InputBlaster~\cite{liu2023testinglimitsunusualtext} employs LLMs to generate specialized text inputs for mobile applications, significantly increasing bug detection rates compared to traditional methods. Similarly, ChatAFL~\cite{meng2024large} and ChatFuzz~\cite{hu2023augmenting} integrate LLMs to enhance the fuzzing of network protocols and web applications, resulting in higher code coverage and vulnerability detection rates. 
Another important aspect of LLM-based fuzzing is the mutation of seed inputs. In traditional fuzzing, mutation strategies such as bit-flipping and block-based mutations are used to generate new test cases~\cite{wang2019superion}. However, LLM-based fuzzing can perform more sophisticated mutations by understanding the context of the input, making them particularly effective for generating valid yet diverse test cases. For example, CovRL-Fuzz uses LLMs to perform mutation operations, guided by coverage feedback, ensuring that mutations not only maintain grammatical correctness but also increase coverage, thus making it more likely to trigger unexpected software behavior~\cite{10.1145/3650212.3680389}.

In addition to improving traditional fuzzing techniques, LLMs have increasingly been employed to automate the generation of fuzz drivers, which are essential for targeting specific APIs or software functions during the fuzzing process. Zhang et al.~\cite{zhang2023understanding} demonstrate the use of GPT-3.5 and GPT-4 to automatically generate fuzz drivers for complex library APIs, achieving an automation rate of over 60\%. This automation significantly reduces the manual effort required to create fuzz drivers and helps streamline the fuzz driver generation process, which is typically a bottleneck in fuzzing campaigns, particularly for large and complex software systems. Similarly, Lyu et al.~\cite{lyu2023prompt} propose PromptFuzz, a coverage-guided fuzzer that leverages LLMs to iteratively generate fuzz drivers capable of exploring previously undiscovered library code.

\section{Methodology}\label{sec:design}
\subsection{Overview}

\begin{figure*}[htbp]
    \centering
    \includegraphics[width=\textwidth]{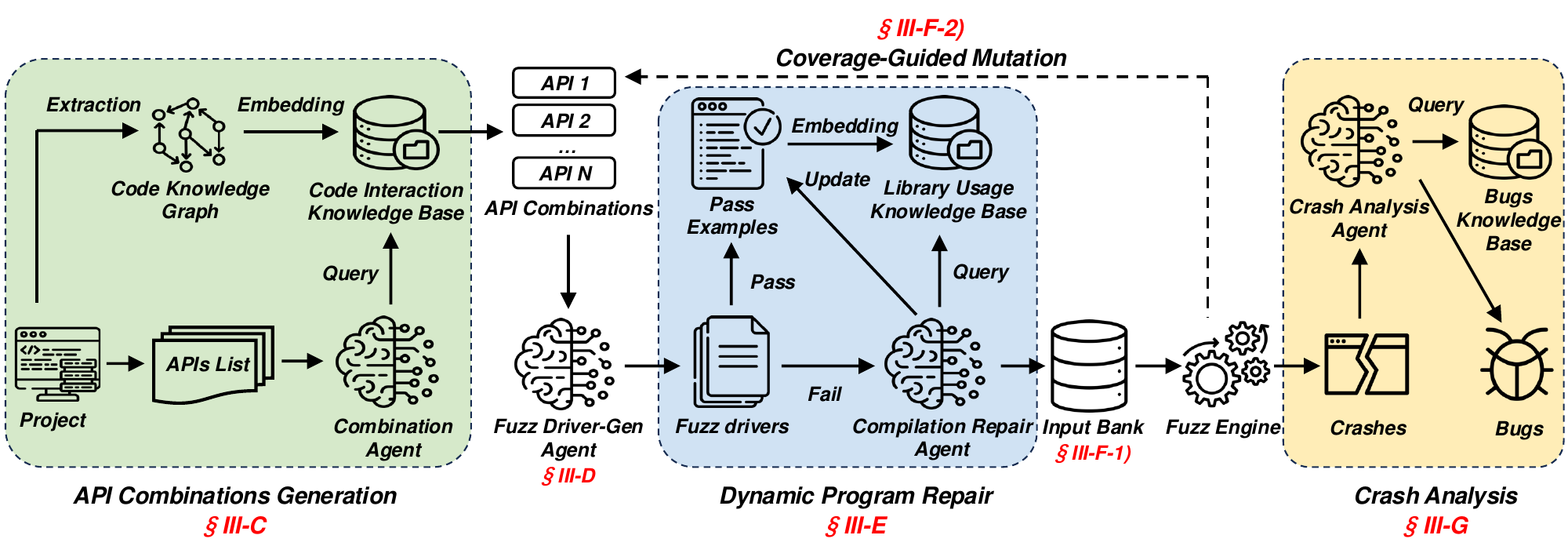}
    \caption{Overview of \approach}
    \label{fig:workflow}
\end{figure*}

In this section, we present the system design of \approach, a fuzz driver generation framework that integrates a multi-agent system with a code knowledge graph. \approach is designed to enhance the quality and coverage of fuzz testing by efficiently generating fuzz drivers for API combinations. The primary objective of our framework is to harness the rich and relative information within the code knowledge graph to capture the complex invocation relationships among library APIs and, based on this, generate highly effective fuzz drivers. By introducing a coverage-guided mutation strategy, \approach continuously optimizes API combinations, enabling the exploration of additional code paths and the identification of hidden vulnerabilities. \autoref{fig:workflow} illustrates the workflow of \approach, which comprises three main components and a unique feedback loop mechanism.

Initially, \approach parses the target project and its library APIs, extracting and embedding a code knowledge graph from the project under test. This parsing process involves two steps: first, parsing the abstract syntax tree, and second, performing interprocedural program analysis. The parser extracts key information such as data structures, function implementations, function signatures, and calling relationships.
Next, \approach queries API combinations for each library API, focusing on those that have invocation relationships or functional similarities, and generates corresponding fuzz drivers. \approach then attempts to compile these generated fuzz drivers and resolves the compilation errors that arise. During the driver generation and repairing, we provide \approach with a dynamically updated knowledge base on library usage.
Once the fuzz drivers are successfully compiled, \approach executes them while monitoring the code coverage of each library file. It employs a code coverage-guided iterative process to mutate API combinations that fail to cover new paths. This iterative process continues until either new code paths are discovered or the mutation budget is exhausted.
Finally, \approach uses chain-of-thought reasoning to analyze any crashes produced during fuzzing. To verify the validity of these crashes, we reference an LLM-generated knowledge base containing real CWE-related source code examples.

\subsection{Construction of the Code Knowledge Graph}
We begin by employing a syntax parser and static analysis tool to extract detailed information on functions, file structures, and function call relationships from the code repository. In addition, we gather API function details and relevant documentation from existing repository resources. To further enhance understanding, we utilize an LLM to generate concise code summaries, usage scenarios, and file-level overviews.

We now introduce the symbols used to describe the code knowledge graph. Let the set of functions in the codebase be 
\[
  \mathcal{F} = \{ f_1, f_2, \dots, f_N \},
\]
where each function \( f_i \) has the following attributes: source code \( c_{f_i} \), signature \( \sigma_{f_i} \), associated file \( d_{f_i} \), and function summary \( s_{f_i} \).

The set of files in the codebase is 
\[
  \mathcal{D} = \{ d_1, d_2, \dots, d_M \},
\]
where each file \( d_j \) have a summary \( s_{d_j} \).

The extracted call relationships are denoted as 
\[
  \mathcal{C} = \{ (f_i, f_j) \mid f_i \text{ calls } f_j \}.
\]
We also define a mapping from functions to their summaries as 
\[
  S: f_i \mapsto s_{f_i}.
\]

The construction process proceeds as follows:
\paragraph{Initialize Knowledge Graph \( G \)} 
We first create an empty property graph \( G = (V, E) \), where \( V \) represents the set of nodes and \( E \) represents the set of edges.

\paragraph{Create Function Nodes} 
For each function \( f_i \in \mathcal{F} \), we create a function node \( v_{f_i} \in V \), with attributes including its signature \( \sigma_{f_i} \), file path \( d_{f_i} \), source code \( c_{f_i} \), and summary \( s_{f_i} \).
\paragraph{Create File Nodes}  
For each file \( d_j \in \mathcal{D} \), we create a file node \( v_{d_j} \in V \) with attributes such as the file summary \( s_{d_j} \). For each function \( f_i \in \mathcal{F} \) where \( d_{f_i} = d_j \), we create an edge \( e_{d_j,f_i} = (v_{d_j}, v_{f_i}) \in E \) labeled as ``CONTAINS''.
\paragraph{Create Call Relationship Edges}  
For each call relationship \( (f_i, f_j) \in \mathcal{C} \), if both \( f_i \) and \( f_j \) belong to \( \mathcal{F} \), we create an edge \( e_{f_i,f_j} = (v_{f_i}, v_{f_j}) \in E \) labeled as ``CALLS''. If \( f_j \) is an external library function, we create an external function node \( v_{f_j} \), labeled as ``LIBRARY\_FUNCTION'', and create an edge \( e_{f_i,f_j} = (v_{f_i}, v_{f_j}) \in E \) labeled as ``LIBRARY\_CALLS''.

\paragraph{Build Property Graph Index} 
We utilize multiple forms of graph indexing, including code indexing and natural language indexing. Since code and natural language reside in different embedding spaces, relying on one type of index for queries of the other type may lead to inaccurate retrieval. Therefore, we adopt separate indexes for each data format: natural language and code. We create two types of property graph indexes: \( PGI_{NL} = (G, \mathcal{V}_{NL}) \) for natural language and \( PGI_{code} = (G, \mathcal{V}_{code}) \) for code, ensuring efficient querying and retrieval.

\subsection{API Combination Generation}
The selection of API combinations is a critical step in guiding the generation of fuzz drivers, as cleverly combining APIs within a single driver can reveal more unexplored execution paths within the target library. \approach employs an external retrieval-augmented process, combining code knowledge graphs with LLM to generate optimal API combinations, which are then used to create fuzz drivers that maximize code coverage. As shown in \autoref{fig:workflow}, the \approach framework follows a multi-step method for API combination generation.

This generation process involves querying the code knowledge graph to retrieve relevant API combinations based on a target API. 
The query process is divided into three key stages: Indexing, Retrieval, and Response Synthesis, as illustrated in \autoref{query}.

\begin{algorithm}
\caption{Query for API Combinations}
\label{query}
\SetKwInput{KwInput}{Input}
\SetKwInput{KwOutput}{Output}
\SetKwFunction{FEmbed}{Embed}
\SetKwFunction{FRetrieve}{RetrieveChunks}
\SetKwFunction{FLLM}{LLM}
\SetKwFunction{FRefine}{RefineWithLLM}

\KwInput{Query $Q$, Knowledge graph $G$, Embedding model $M$, Similarity threshold $s$, Number of chunks $k$, LLM $L$, Refine prompt $P_r$}
\KwOutput{API combination $C$}

$IndexBase \leftarrow \FEmbed(G, M)$\;

$Q_{vec} \leftarrow \FEmbed(Q, M)$\;

$C_1, C_2, \dots, C_k \leftarrow \FRetrieve(Q_{vec}, IndexBase, s, k)$\;

$R \leftarrow \FLLM(Q, C_1)$\;
\ForEach{remaining chunk $C_i$ in $C_2, \dots, C_k$}{
    $R \leftarrow \FRefine(R, C_i, P_r, L)$\;
}

$C \leftarrow \FLLM(Q, R)$\;

\Return $C$\;
\end{algorithm}

In the indexing phase, the knowledge graph is transformed into an index base by splitting it into smaller, manageable chunks. Each chunk represents a subset of the graph's content, allowing the system to efficiently retrieve relevant information during the query process. The indexing ensures that the knowledge graph is well-organized for querying by the LLM.

When the LLM issues a query, representing a request for APIs related to a target API, the system retrieves a set number of chunks from the indexed knowledge graph. The selection of these chunks is based on their relevance to the query, ensuring that the most pertinent information is retrieved to form a response.

The retrieved chunks are processed sequentially, with the final response synthesized through an iterative refinement process~\cite{madaan2024self}. The initial chunk initializes the query answer, prompting the LLM to iteratively incorporate the useful remaining chunks while filtering out noisy data. This optimization continues until all retrieved information is considered. The final output is a set of APIs functionally related to the target API. Through iterative refinement, all useful information is integrated into the final response, minimizing the impact of noisy data.

\subsection{Fuzz Driver Generation}

To generate high-quality fuzz drivers, we adopt a role-prompt strategy combined with a memory mechanism to guide the LLMs~\cite{kong2024betterzeroshotreasoningroleplay,kwon2023efficient}. These prompts are constructed using both the source code and natural language descriptions of the API combinations, ensuring that the LLM generates fuzz drivers that are not only syntactically correct but also contextually aligned with the APIs’ functionality.

Our prompt strategy emphasizes several key aspects of fuzz driver generation:
\begin{itemize}
    \item \textbf{Task Definition}: The LLM is tasked with generating a fuzz driver that tests the provided API combination. The prompt specifies that each API must be invoked within the function \texttt{LLVMFuzzerTestOneInput}, ensuring comprehensive testing coverage of the API set.
    \item \textbf{API Context}: The prompt includes the API source code, headers, and natural language summaries. This provides the LLM with the necessary context to correctly utilize the APIs and manage their inputs and outputs according to their functionality and constraints.
    \item \textbf{Error Handling}: To prevent instability in the fuzz driver itself, which could lead to crashes and reduce overall fuzzing efficiency, the prompt includes explicit instructions for robust error handling and careful memory management. 
\end{itemize}

By constructing the prompt with these components, we ensure that the LLM generates fuzz drivers that are both executable and aligned with the best practices of fuzz testing. The use of the memory mechanism allows the LLM to refine its generated response iteratively, improving the quality of the fuzz driver as it processes additional information from the API combination.

\subsection{Dynamic Program Repair}
Due to hallucinations and a lack of high-quality training data, code generated by LLMs often contains syntax and semantic errors~\cite{liu2023codegeneratedchatgptreally,pearce2021asleepkeyboardassessingsecurity,qiu2024efficientllmgeneratedcoderigorous}, which may affect the reliability of fuzz drivers. A robust fuzz driver must ensure correctness at the code level; however, we observed that some fuzz drivers produced by \approach fail to compile. These failures are primarily caused by syntax errors and incorrect usage of library APIs. To address this issue, we developed an LLM-based dynamic program repair mechanism that automatically corrects these errors and improves the overall stability of fuzz drivers, as shown in \autoref{fix}. Compared to traditional static rule analysis, LLM-based solutions excel at handling complex, context-dependent errors that are difficult to resolve through static rules alone~\cite{6100080,10.1145/3611643.3613892,joshi2022repairnearlygenerationmultilingual}.

\begin{algorithm}[]
\caption{Fuzz Driver Repair}
\label{fix}
\SetAlgoLined
\DontPrintSemicolon
\SetKwInput{KwInput}{Input}
\SetKwInput{KwOutput}{Output}
\SetKwFunction{FInitKB}{InitKB}
\SetKwFunction{FCompile}{Compile}
\SetKwFunction{FQueryKB}{QueryKB}
\SetKwFunction{FUpdateKB}{UpdateKB}
\SetKwFunction{FLLMRepair}{LLMRepair}
\SetKwFunction{FConstructQuery}{ConstructQuery}

\KwInput{Fuzz driver list $D_{list}$, API usage samples $S$, Max iterations $M$, LLM $L$}
\KwOutput{Repaired drivers $D_{rep}$, Knowledge base $KB$}

$D_{rep} \leftarrow \emptyset$\; $KB \leftarrow \FInitKB(S)$; 

\ForEach{$D \in D_{list}$}{
    $i \leftarrow 0$; $compiled \leftarrow false$\;
    \While{$i < M$ \textbf{and} not $compiled$}{
        $result, errors \leftarrow \FCompile(D)$\;
        \uIf{$result$ \textbf{is} success}{
            $compiled \leftarrow true$; $D_{rep} \leftarrow D_{rep} \cup \{D\}$\;
            $KB \leftarrow \FUpdateKB(KB, D)$\;
        }
        \Else{
                $query \leftarrow \FConstructQuery(errors)$\;
                $correct\_cases \leftarrow \FQueryKB(L, KB, query)$\;
                $D \leftarrow \FLLMRepair(L, D, correct\_cases)$\;
            
        }
        $i \leftarrow i + 1$\;
    }
}
\Return{$D_{rep}, KB$}
\end{algorithm}

To implement dynamic program repair, we first initialize a knowledge base containing correct library API usages, constructed from samples of fuzz drivers from OSS-Fuzz and the library’s header files. It should be noted that while the OSS-Fuzz code focuses on individual APIs, our approach handles API combinations. The header files help resolve errors caused by missing include statements.
When \approach encounters a compilation failure, it processes the compiler error messages and constructs a query to search the external knowledge base for the correct usage of the erroneous API or code snippet. \approach continues to query the knowledge base and apply fixes iteratively until the fuzz driver successfully compiles or a maximum iteration limit is reached.

Additionally, all successfully compiled fuzz drivers are inserted back into the knowledge base, dynamically updating the repository of correct API usage patterns. This dynamic update strategy ensures that the knowledge base evolves over time, gradually expanding to cover a wider range of API usage scenarios across different libraries. As a result, our approach maintains a high level of adaptability and accuracy in fixing errors, reducing the chances of recurrent compilation failures.

\subsection{Fuzzing Loop}

The fuzzing loop in \approach is designed to iteratively improve the effectiveness of fuzz testing by initializing high-quality inputs and guiding mutations based on coverage feedback. The loop consists of two main components: the input bank and the coverage-guided mutation mechanism, both working together to maximize code coverage and identify new execution paths.

\subsubsection{Input bank initialization}

Before fuzzing begins, we initialize an input bank by generating input seeds for each fuzz driver using LLMs. The input generation process starts with a detailed analysis of the fuzz driver's data flow, where we extract the value flow relationships between variables. Subsequently, we extract the API function signatures that the fuzz driver interacts with, which helps in understanding the structure and constraints of the expected inputs.

Using this information, we construct a prompt for the LLM, instructing it to generate input seeds that maximize code coverage, target edge cases, and explore boundary conditions. The generated inputs are designed to adhere to the data flow dependencies and the API’s input requirements, ensuring that they are both valid and effective in triggering diverse execution paths from the very outset of fuzzing.

The initial input seeds stored in the input bank serve as a foundation for the fuzzing process. As fuzzing progresses, the input bank is continuously updated with new inputs generated from coverage feedback, ensuring that the fuzzing loop remains dynamic and adaptive.

\subsubsection{Coverage-guided mutation}

During the fuzzing process, \approach monitors the code coverage achieved by each fuzz driver. It analyzes the coverage data across the library and identifies files with lower coverage compared to the overall library. For each low-coverage file, the defined API functions are extracted, and a prioritized list, known as the low coverage API list, is generated based on the coverage rankings of the files.

\begin{algorithm}[]
\caption{Coverage-Guided Mutation}
\label{mutation}
\SetKwInput{KwInput}{Input}
\SetKwInput{KwOutput}{Output}
\SetKwFunction{FAnalyze}{AnalyzeFileCoverage}
\SetKwFunction{FExtract}{ExtractAPIInfo}
\SetKwFunction{FSort}{SortByCoverage}
\SetKwFunction{FMutate}{MutateAPICombination}
\SetKwFunction{FDetect}{DetectNewPath}
\SetKwFunction{FConstructQuery}{ConstructQuery}

\KwInput{Library $F$, Current API combination $A$, Current library coverage $C_{current}$, LLM $L$, Max iterations $M$}
\KwOutput{Updated API combination $A_{new}$}

$lowCovFiles \leftarrow \{\}$\;
$lowCovAPIs \leftarrow \{\}$\;

\ForEach{$file \in F$}{
    $fileCoverage \leftarrow \FAnalyze(file, A)$\;
    \If{$fileCoverage < C_{current}$}{
        $lowCovFiles \leftarrow lowCovFiles \cup \{file\}$\;
        $apis \leftarrow \FExtract(file)$\;
        $lowCovAPIs \leftarrow lowCovAPIs \cup apis$\;
    }
}

$sortedLowCovAPIs \leftarrow \FSort(lowCovAPIs, lowCovFiles)$\;
$i \leftarrow 0$\;
$if\_newPath \leftarrow false$\;

\While{$i < M$ \textbf{and} not $newBranch$}{
    $query \leftarrow  \FConstructQuery(sortedLowCovAPIs)$\;
    $A_{new} \leftarrow \FMutate(L, A, query)$\;
    $if\_newPath \leftarrow \FDetect(F, A_{new})$\;
    $i \leftarrow i + 1$\;
}

\Return{$A_{new}$}
\end{algorithm}

This low coverage API list is then used to guide the LLM in mutating and restructuring the current API combinations in the fuzz driver. The objective is to explore new execution paths and increase code coverage by focusing on under-explored APIs. As shown in \autoref{mutation}, the mutation process is iterative: \approach continues to query the LLM with new API combinations based on the low coverage API list until new branches are covered or computational resources are exhausted.

These mutation strategies are guided by the coverage feedback from previous executions. By prioritizing low-coverage APIs, \approach focuses on areas of the code that are less explored, maximizing the likelihood of discovering new program behaviors.

\subsection{Crash Analysis}

During the fuzzing process, \approach is equipped with several runtime error detection tools, like Address-Sanitizer~\cite{10.5555/2342821.2342849} (ASan), Undefined-Behavior-Sanitizer~\cite{Undefinedbehaviorsanitizer} (UBSan) and MemorySanitizer (MSan)~\cite{7054186}. However, the crashes detected by these tools can originate from different sources, including incorrect API usage in the fuzz driver or bugs within the library APIs themselves. To distinguish between these causes, we developed an automatic crash analysis module designed to identify the root cause of each crash.

The crash analysis module employs a chain-of-thought strategy, guiding the LLM through a structured, step-by-step reasoning process to diagnose the source of each crash. This methodical approach ensures that each stage builds upon the insights gained from the previous one, leading to a clear understanding of the crash’s root cause. The process unfolds through the following key steps:

\begin{itemize}
    \item \textbf{Step 1: Source Code Extraction}: The analysis begins by isolating the specific code regions where the crash occurred. This includes extracting relevant portions of the fuzz driver’s source code, along with any associated API calls that may have influenced the crash.

    \item \textbf{Step 2: Error Pattern Hypothesis}: After extracting the relevant code, the LLM systematically analyzes the code for patterns indicative of common programming errors. These patterns could include unsafe memory operations, incorrect variable assignments, improper control flow conditions and so on. At this stage, the LLM forms initial hypotheses about which code patterns might be responsible for triggering the crash, setting up a list of plausible error sources for deeper exploration.
    
    \item \textbf{Step 3: CWE-Based Pattern Matching}: Armed with hypotheses about potential error patterns, the LLM constructs targeted queries to search a CWE-based knowledge base. This knowledge base, comprising over 100 real-world CWE vulnerabilities specific to C/C++ programs, provides a wealth of information about known issues. 
\end{itemize}

By progressively narrowing down the potential causes and comparing them against known vulnerabilities, the chain-of-thought strategy ensures that the LLM can systematically trace the crash to its most likely source, whether it stems from fuzz driver errors or underlying library bugs.

The knowledge base used in the crash analysis module is constructed from a collection of real-world CWE vulnerabilities specific to C and C++ programs. Each entry in the knowledge base includes:
\begin{itemize}
    \item A natural language description of the vulnerability, detailing its characteristics and typical causes.
    \item A corresponding code sample generated by the LLM based on the vulnerability description, exemplifying the vulnerability in a real-world context.
\end{itemize}

The natural language descriptions serve as prompts for the LLM to better understand the crash context. The inclusion of example code allows the LLM to draw comparisons between the source code of fuzz driver or library APIs and known vulnerabilities, improving the accuracy of the crash analysis.

\section{Evaluation}

\begin{table*}[htbp]
\caption{Code coverage result of fuzz drivers generated by \approach}
\centering
\setlength{\tabcolsep}{0pt}
\begin{tabular*}{\textwidth}{@{\extracolsep{\fill}}lllccccc@{}}
\toprule
\multirow{2}{*}{\textbf{Library}} & \multirow{2}{*}{\textbf{Version}} & \multirow{2}{*}{\textbf{Branches}} & \multirow{2}{*}{\textbf{APIs}} & \multirow{2}{*}{\makecell{\textbf{Generated Fuzz drivers}}} & \multicolumn{3}{c}{\textbf{Branch Coverage}} \\
\cmidrule(l){6-8} 
 &  &  &  &  & \textbf{\approach{}} & \textbf{PromptFuzz} & \textbf{OSS-Fuzz} \\
\midrule
c-ares & 1.8 & 6,390 & 119 & 101 & \textbf{54.99\%(3,514)} & 48.73\%(3,114) & 18.79\%(1,201)\\
cjson & 1.7.18 & 1,379 & 41 & 41 & \textbf{84.48\%(1,165)} &74.11\%(1,022) &38.07\%(525) \\
curl & 8.9 & 19,501 & 92 & 88 & \textbf{26.68\%(5,203)} & 19.73\%(3,847) & 4.67\%(911)\\
lcms & 2.15 & 7,482 & 216 & 205 & \textbf{46.93\%(4,693)} & 40.71\%(3,046) & 39.79\%(2,977) \\
libpcap & 1.10.4 & 5,227 & 67 & 64 & \textbf{46.80\%(2,446)} & 39.53\%(2,066) & 40.81\%(2,133) \\
libtiff & 4.7 &16,347 & 169 & 161 &38.81\%(6,344) &\textbf{48.29\%(7,894)} &38.83\%(6,348) \\
libvpx & 1.14 &17,640 &31 & 29 &\textbf{22.10\%(3,899)} &21.09\%(3,721) & 15.50\%(2,734) \\
zlib & 1.3 & 2,375 & 63 & 61 & 61.6\%(1,463) & \textbf{72.04\%(1,711)} &51.62\%(1,226) \\
\bottomrule
\end{tabular*}
\label{tab:code_cov}
\end{table*}

\begin{table*}
\caption{Comparison of different variants across multiple libraries}
\setlength{\tabcolsep}{0pt}
\fontsize{6}{7}\selectfont
\begin{tabular*}{\textwidth}{@{\extracolsep{\fill}}l*{16}{c}}
\toprule
\multirow{2}{*}[-0.5ex]{\textbf{Variants}} & \multicolumn{2}{c}{\textbf{c-ares}} & \multicolumn{2}{c}{\textbf{cjson}} & \multicolumn{2}{c}{\textbf{curl}} & \multicolumn{2}{c}{\textbf{lcms}} & \multicolumn{2}{c}{\textbf{libpcap}} & \multicolumn{2}{c}{\textbf{libtiff}} & \multicolumn{2}{c}{\textbf{libvpx}} & \multicolumn{2}{c}{\textbf{zlib}} \\
\cmidrule(lr){2-3} \cmidrule(lr){4-5} \cmidrule(lr){6-7} \cmidrule(lr){8-9} \cmidrule(lr){10-11} \cmidrule(lr){12-13} \cmidrule(lr){14-15} \cmidrule(lr){16-17}
& \makecell{Pass\\ Rate} & Cov. & \makecell{Pass\\ Rate} & Cov. & \makecell{Pass\\ Rate} & Cov. & \makecell{Pass\\ Rate} & Cov. & \makecell{Pass\\ Rate} & Cov. & \makecell{Pass\\ Rate} & Cov. & \makecell{Pass\\ Rate} & Cov. & \makecell{Pass\\ Rate} & Cov. \\
\midrule
Without repair & 46.22\% & 27.14\% & 82.93\% & 71.57\% & 69.57\% & 22.44\% & 56.94\% & 36.70\% & 58.21\% & 35.28\% & 64.5\% & 28.01\% & 22.58\% & 9.31\% & 42.86\% & 41.52\% \\
 & (55/119) & (1,734) & (34/41) & (987) & (64/92) & (4,376) & (123/216) & (2,746) & (39/67) & (1,844) & (109/169) & (4,586)& (7/31) & (1,642) & (27/63) & (986) \\
\addlinespace
LLM-only repair & 73.96\% & 44.87\% & 97.56\% & 77.08\% & 85.87\% & 23.46\% & 72.22\% & 43.50\% & 73.13\% & 39.93\% & 79.29\% & 35.32\% & 51.61\% & 16.63\% & 85.71\% & 55.12\% \\
 & (88/119) & (2,867) & (40/41) & (1,063) & (79/92) & (4,574) & (156/216) & (3,255) & (49/67) & (2,087) & (134/169) & (5,773) & (16/31) & (2,934) & (54/63) & (1,309) \\
\midrule
Text-only retrieval & - & 47.17\% & - & 79.70\% & - & 22.09\% & - & 40.68\% & - & 40.16\% & - & 36.73\% & - & 21.22\% & - & 55.79\% \\
 & & (3,014) & & (1,099) & & (4,307) & & (3,044) & & (2,099) & & (6,005) & & (3,744) & & (1,325) \\
\midrule
\approach & \textbf{84.87\%} & \textbf{54.99\%} & \textbf{100\%} & \textbf{84.48\%} & \textbf{95.65\%} & \textbf{26.68\%} & \textbf{94.91\%} & \textbf{46.93\%} & \textbf{95.52\%} & \textbf{46.80\%} & \textbf{94.08\%} &\textbf{38.81\%} & \textbf{93.55\%} & \textbf{22.10\%} & \textbf{96.83\%} & \textbf{61.6\%} \\
 & \textbf{(101/119)} & \textbf{(3514)} & \textbf{(41/41)} & \textbf{(1,165)} & \textbf{(88/92)} & \textbf{(5,203)} & \textbf{(205/216)} & \textbf{(4,693)} & \textbf{(64/67)} & \textbf{(2,446)} & \textbf{(159/169)} &\textbf{(6,344)} & \textbf{(29/31)} & \textbf{(3,899)} & \textbf{(61/63)} & \textbf{(1,463)} \\
\bottomrule
\end{tabular*}
\label{tab:comparison}
\end{table*}

To evaluate the performance of \approach, we defined three research questions:

\textbf{RQ1}: Is \approach more effective than other fuzzers?

\textbf{RQ2}: How does each agent contribute to enhancing the effectiveness of \approach?

\textbf{RQ3}: Can \approach accurately analyze the cause of crashes generated during fuzzing?

\subsection{Experimental Environment}
\subsubsection{Implementation}
We use Tree-sitter\footnote{\url{https://tree-sitter.github.io/tree-sitter/}} to parse the source code, which provides an efficient way to build and update the syntax tree as the source file changes. For deeper analysis of the codebase, including extracting API call relationships and source code details, we leverage CodeQL\footnote{\url{https://codeql.github.com/}}. CodeQL allows us to query the entire repository and retrieve relevant information regarding API usage, forming the basis for the subsequent processing stages.

To represent the extracted code and API calls in an embedding space, we evaluated several embedding models and selected the BAAI/bge-small-en-v1.5 model~\cite{BAAI/bge-small-en-v1.5}, which performed well in our experiments. Our approach uses Llama-Index~\cite{llamaindex} for the integration of the embeddings; however, since Llama-Index was not specifically designed for code graphs, we developed a custom RAG engine to better handle code-related tasks. We chose the DeepSeek-V2-Coder model~\cite{liu2024deepseek} to handle fuzz driver generation and compilation-repair tasks. This model was selected due to its low API pricing and its satisfactory performance in code generation tasks. Additionally, we used the DeepSeek-V2-Chat~\cite{liu2024deepseek} model for other agents.

All experiments were performed on a server equipped with an AMD 64-Core Processor and 1TB of RAM, running a 64-bit version of Ubuntu 22.04 LTS.

\subsubsection{Settings}
To evaluate \approach, we selected eight open-source libraries as our test subjects. We conducted the experiments using the OSS-Fuzz platform~\cite{serebryany2017oss} and libFuzzer as the underlying fuzzing environment. We collected the API lists for each library from their respective documentation or official websites. We configured the temperature parameter for DeepSeek-V2-Coder to 0.7 and for DeepSeek-V2-Chat to 1.0. The maximum length for the default API combinations was set to 6, while the maximum number of iterations for program repair was limited to 5. Additionally, the maximum number of mutations for API combinations was set to 3. To ensure that each library was fuzzed for an adequate period, we assigned different time budgets to the fuzz drivers of each library, making sure that the total fuzzing time for each library reached 24 CPU hours. Every experiment was repeated five times to mitigate statistical errors, and the average results were reported.

\subsection{RQ1. Comparison with existing fuzzers}
To address \textbf{RQ1}, we selected eight open-source libraries for evaluation: \texttt{c-ares}, \texttt{cjson}, \texttt{curl}, \texttt{lcms}, \texttt{libpcap}, \texttt{libtiff}, \texttt{libvpx}, and \texttt{zlib}. These libraries were chosen for their diverse API usage and wide adoption in various software projects. We compared \approach with other open-source fuzzers, focusing on coverage-guided fuzzer (OSS-Fuzz~\cite{serebryany2017oss}) and LLM-based fuzzer (PromptFuzz~\cite{lyu2023prompt}). During the evaluation, we ran the fuzz drivers from OSS-Fuzz on each library for 24 hours. If a library in OSS-Fuzz contained multiple fuzz drivers, we allocated an equal time budget to each driver to ensure the total fuzzing time was 24 hours. For PromptFuzz, we used the same APIs list provided to \approach and controlled the fuzzing duration to be 24 hours as well. We then evaluated the effectiveness of \approach based on code coverage.

The code coverage results for \approach, PromptFuzz, and OSS-Fuzz are shown in \autoref{tab:code_cov}. In comparing \approach with the other fuzzers, \approach demonstrated the highest branch coverage in 6 out of the 8 libraries tested. This superiority can be largely attributed to the structured API combinations generated by \approach, which are derived from a comprehensive code knowledge graph. This allows \approach to capture the complex relationships and semantic dependencies between APIs, thus creating more meaningful fuzz drivers. 

In contrast, PromptFuzz initializes its API combinations through random selections and lacks the deeper understanding of API interdependencies. As a result, \approach consistently outperformed PromptFuzz in libraries that require precise API usage patterns. In the \texttt{lcms} library, \approach covered 46.93\% of branches, while PromptFuzz covered only 40.71\%. These results suggest that \approach's knowledge-driven approach to API combination allows it to explore a larger portion of the codebase, leading to higher coverage. In cases where \approach did not achieve the highest coverage, such as in the \texttt{libtiff} library, the gap between \approach and PromptFuzz was relatively small (38.81\% vs. 48.29\%). This suggests that even when PromptFuzz benefits from its constrained input mechanisms, \approach remains competitive, leveraging its knowledge graph to generate targeted fuzz drivers.

\subsection{RQ2. Ablation study}

In this section, we conduct experiments to investigate the impact of various components of \approach on its effectiveness. The components we evaluate include the code knowledge graph for API combinations generation, compilation repair, and API combination mutation. Detailed experimental results are shown in \autoref{fig:mutation} and \autoref{tab:comparison}.
\subsubsection{Code knowledge graph}
To evaluate the impact of the code knowledge graph on the quality of fuzz drivers generated by \approach, we compared it with a variant that uses text-based API knowledge from libraries. This variant generates API combinations by querying a text-based knowledge base consisting of API source code and API summaries. The experimental results in \autoref{tab:comparison} show that, the text-only retrieval variant achieves lower code coverage compared to \approach, this highlights the effectiveness of the code knowledge graph in generating high-quality fuzz drivers.

\subsubsection{Compilation repair}
In \approach, we introduced a dynamic program repair component to fix fuzz driver programs that encounter compilation errors. In this section, we design experiments to evaluate the effectiveness of \approach in repairing compilation errors and the impact of the dynamic program repair component on fuzzing performance. First, we created two variants: one without any compilation error repair (\texttt{without repair}) and another that relies solely on an LLM for program repair (\texttt{LLM-only repair}). The experimental results in \autoref{tab:comparison} show that the \texttt{without repair} achieved an average compilation success rate of only 57.39\% (458/798), indicating that \approach struggles to directly generate compilable fuzz driver programs. On the other hand, the \texttt{LLM-only repair} increased the compilation success rate to 77.19\% (616/798), demonstrating the significant potential of LLMs for program repair. However, \approach achieved a compilation success rate of 93.99\% (750/798), highlighting the effectiveness of the dynamic program repair component in fixing compilation errors for 
fuzz driver programs.

\begin{figure*}[]
    \centering
    \scalebox{0.95}{ % Scale to 90% of the original size
        \begin{minipage}{\textwidth}
            \centering
            \begin{subfigure}[b]{0.245\textwidth}
                \centering
                \includegraphics[width=\textwidth]{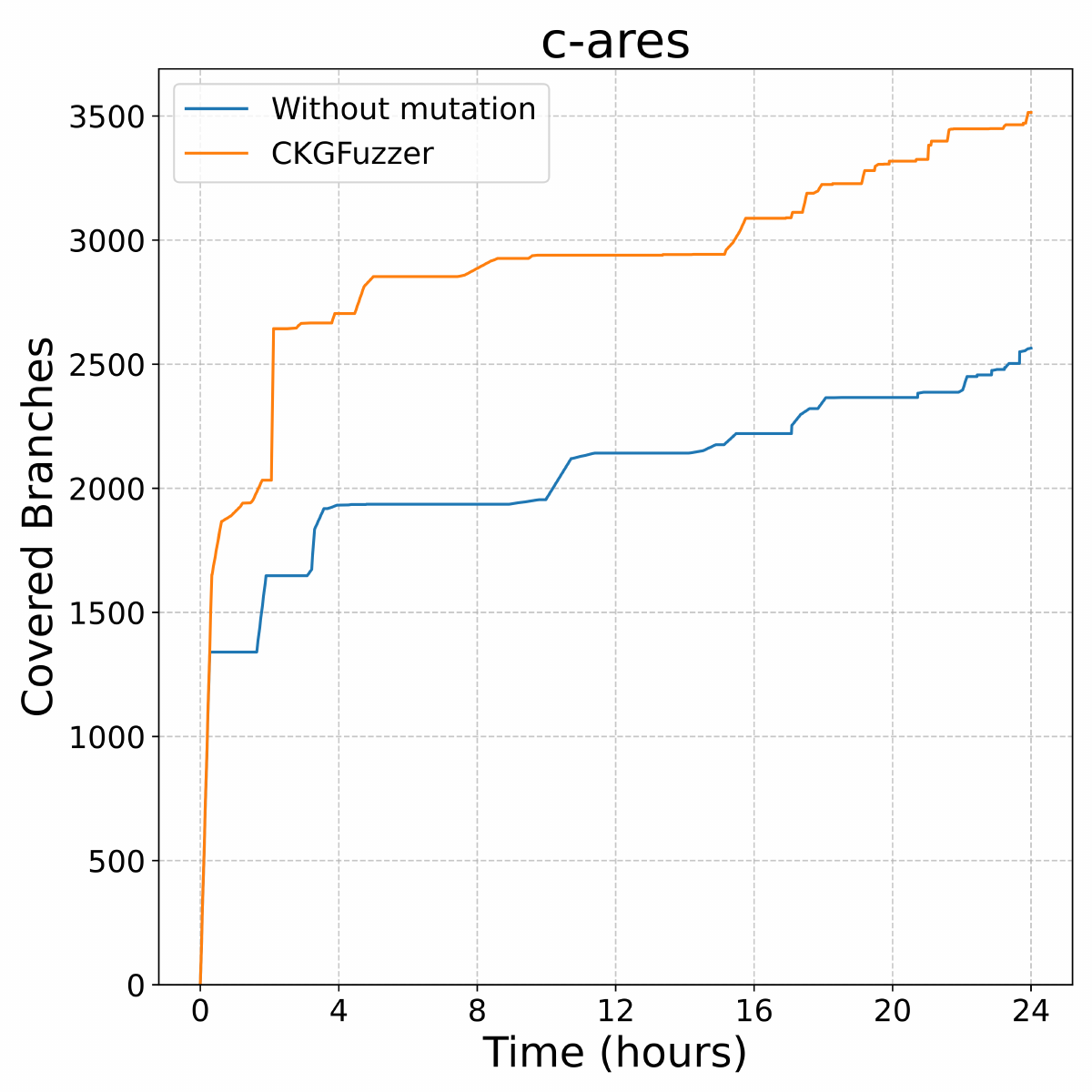}
            \end{subfigure}
            \hfill
            \begin{subfigure}[b]{0.245\textwidth}
                \centering
                \includegraphics[width=\textwidth]{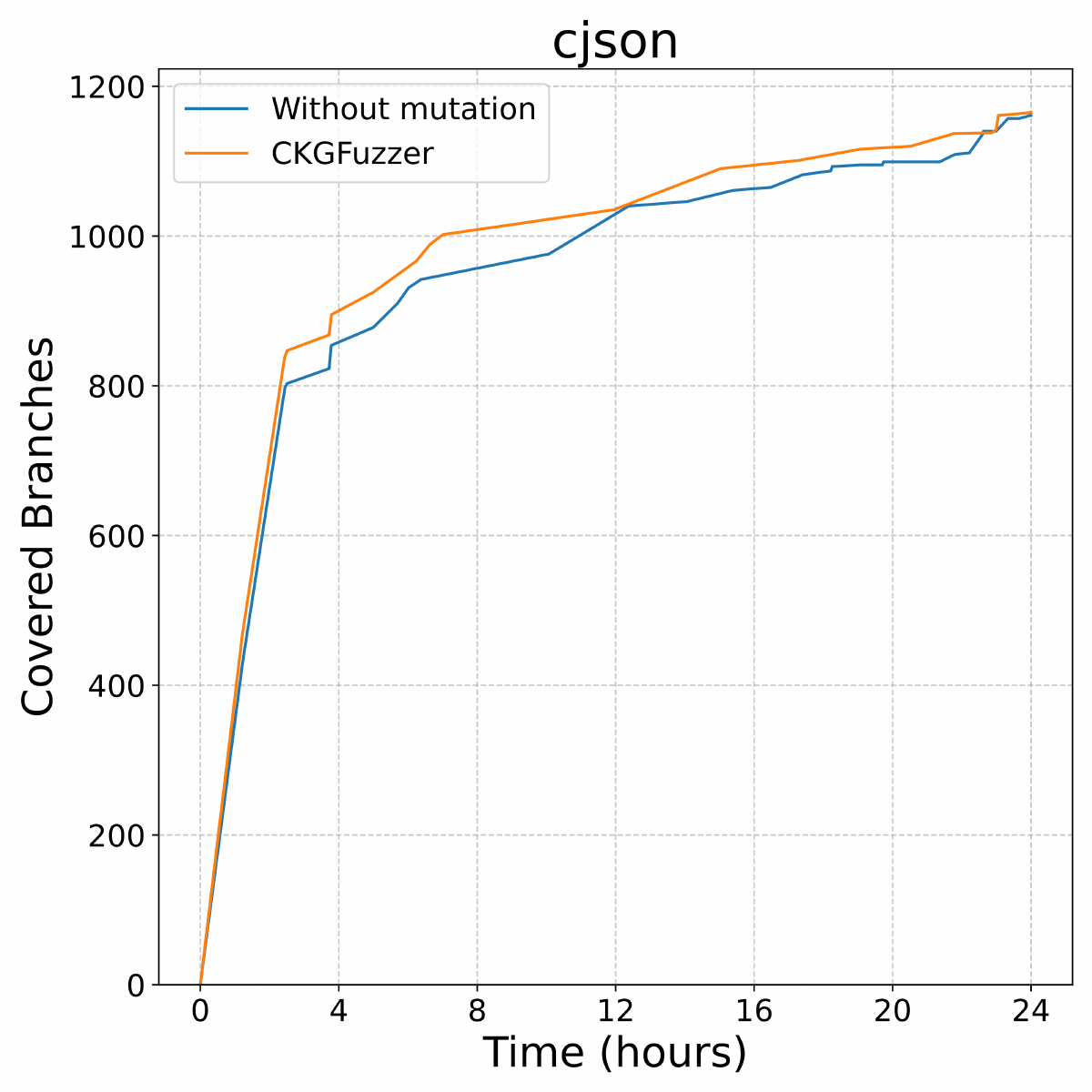}
            \end{subfigure}
            \hfill
            \begin{subfigure}[b]{0.245\textwidth}
                \centering
                \includegraphics[width=\textwidth]{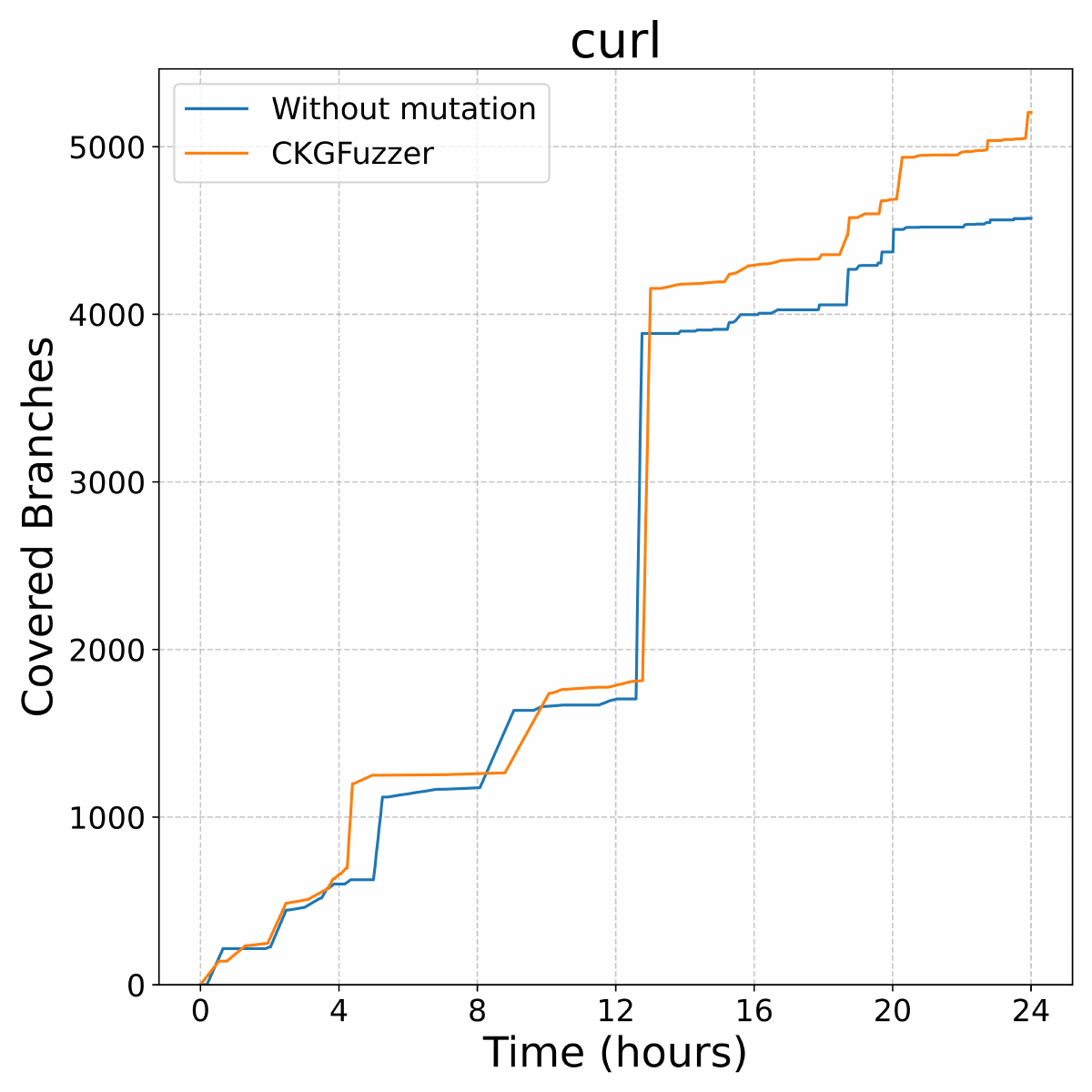}
            \end{subfigure}
            \hfill
            \begin{subfigure}[b]{0.245\textwidth}
                \centering
                \includegraphics[width=\textwidth]{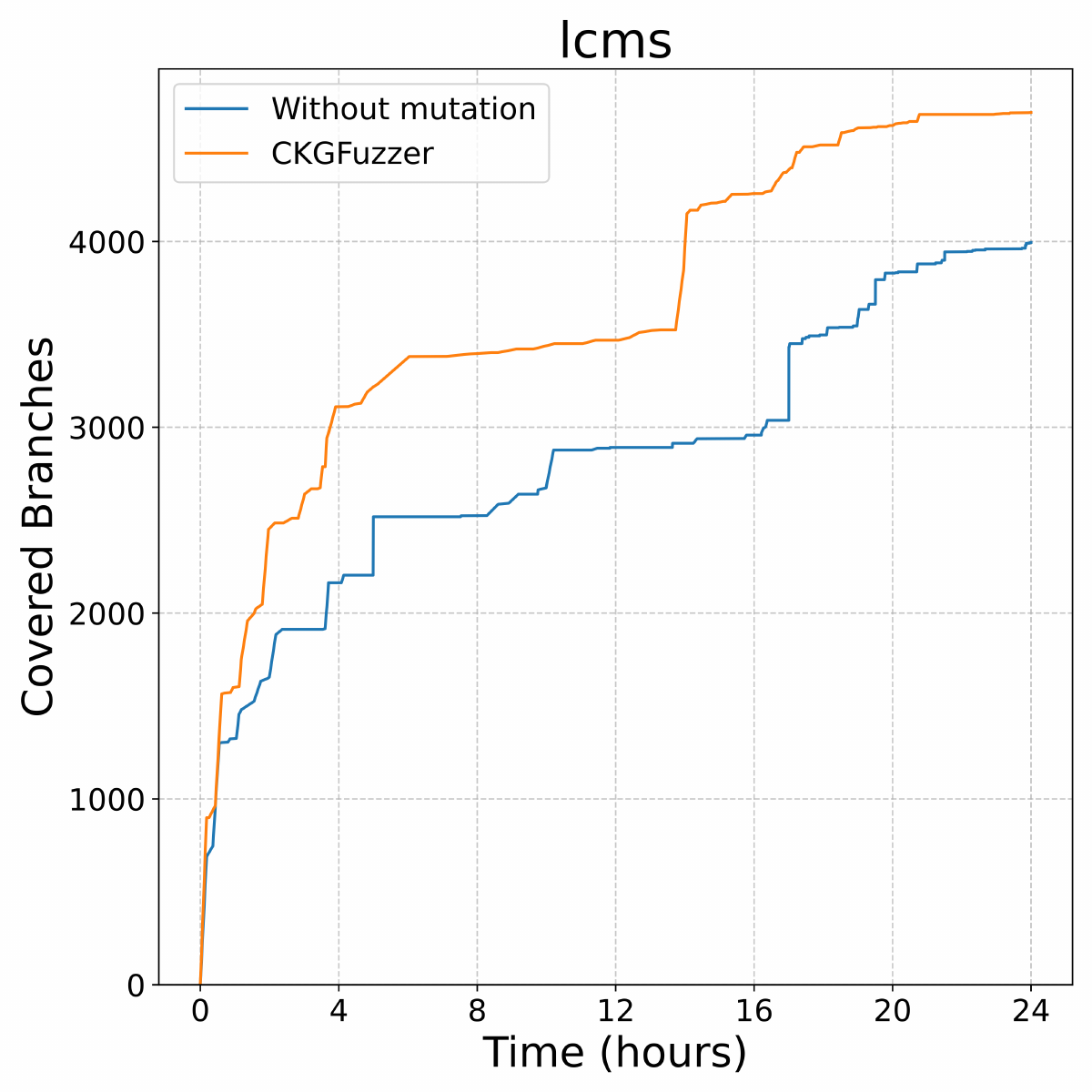}
            \end{subfigure}
        \end{minipage}
    } % End of first row
    \vspace{0.5cm}
    \scalebox{0.95}{ % Second row, also scaled
        \begin{minipage}{\textwidth}
            \centering
            \begin{subfigure}[b]{0.245\textwidth}
                \centering
                \includegraphics[width=\textwidth]{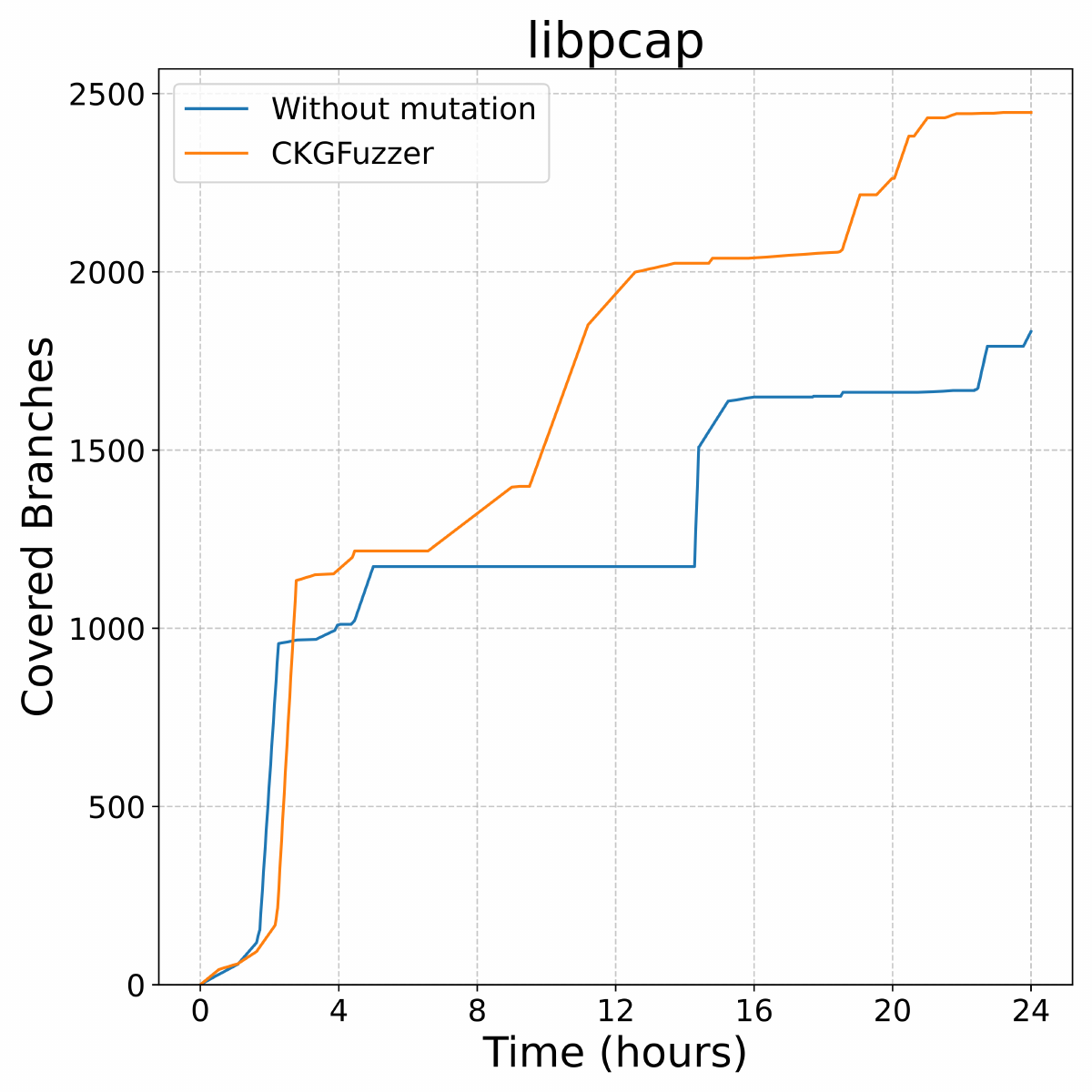}
            \end{subfigure}
            \hfill
            \begin{subfigure}[b]{0.245\textwidth}
                \centering
                \includegraphics[width=\textwidth]{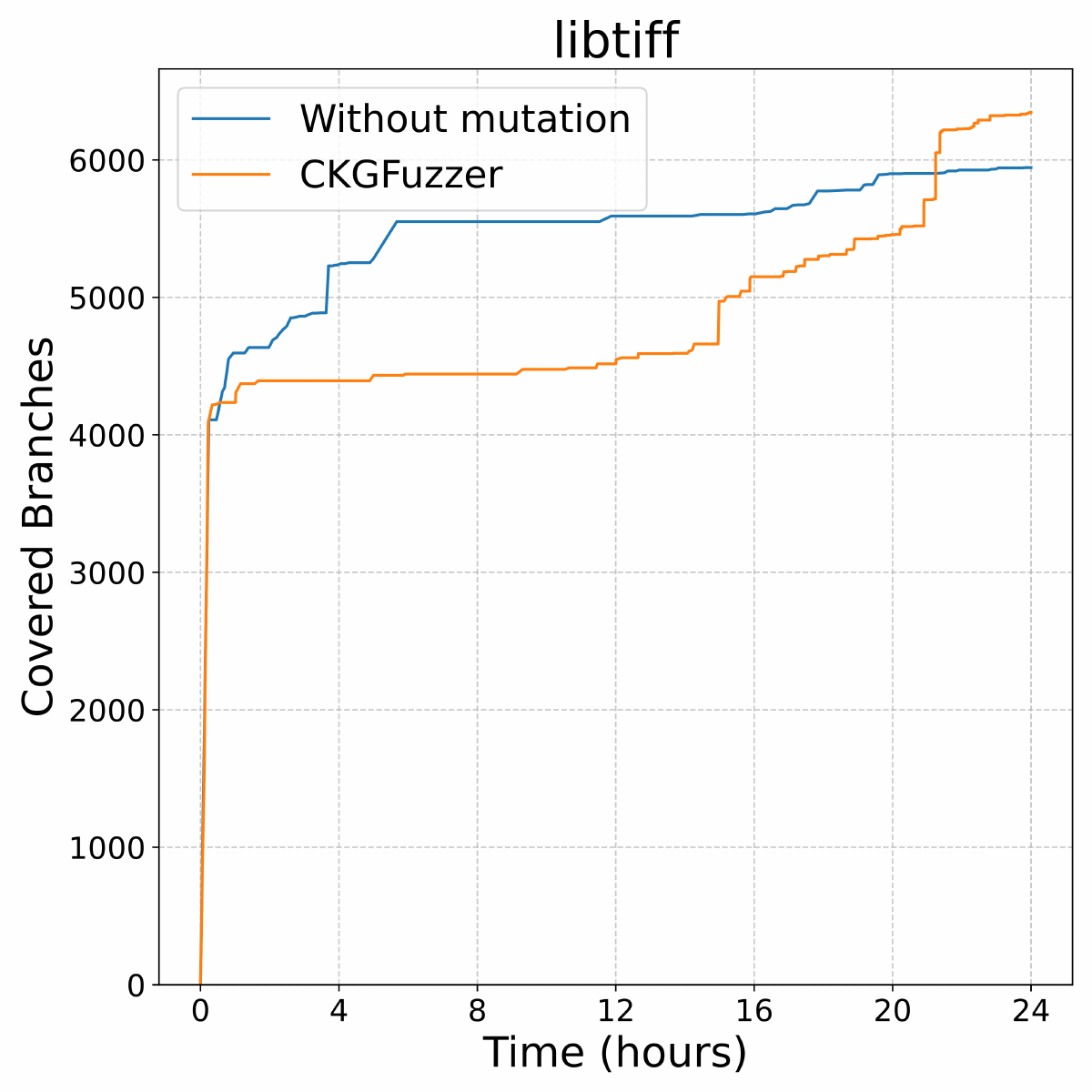}
            \end{subfigure}
            \hfill
            \begin{subfigure}[b]{0.245\textwidth}
                \centering
                \includegraphics[width=\textwidth]{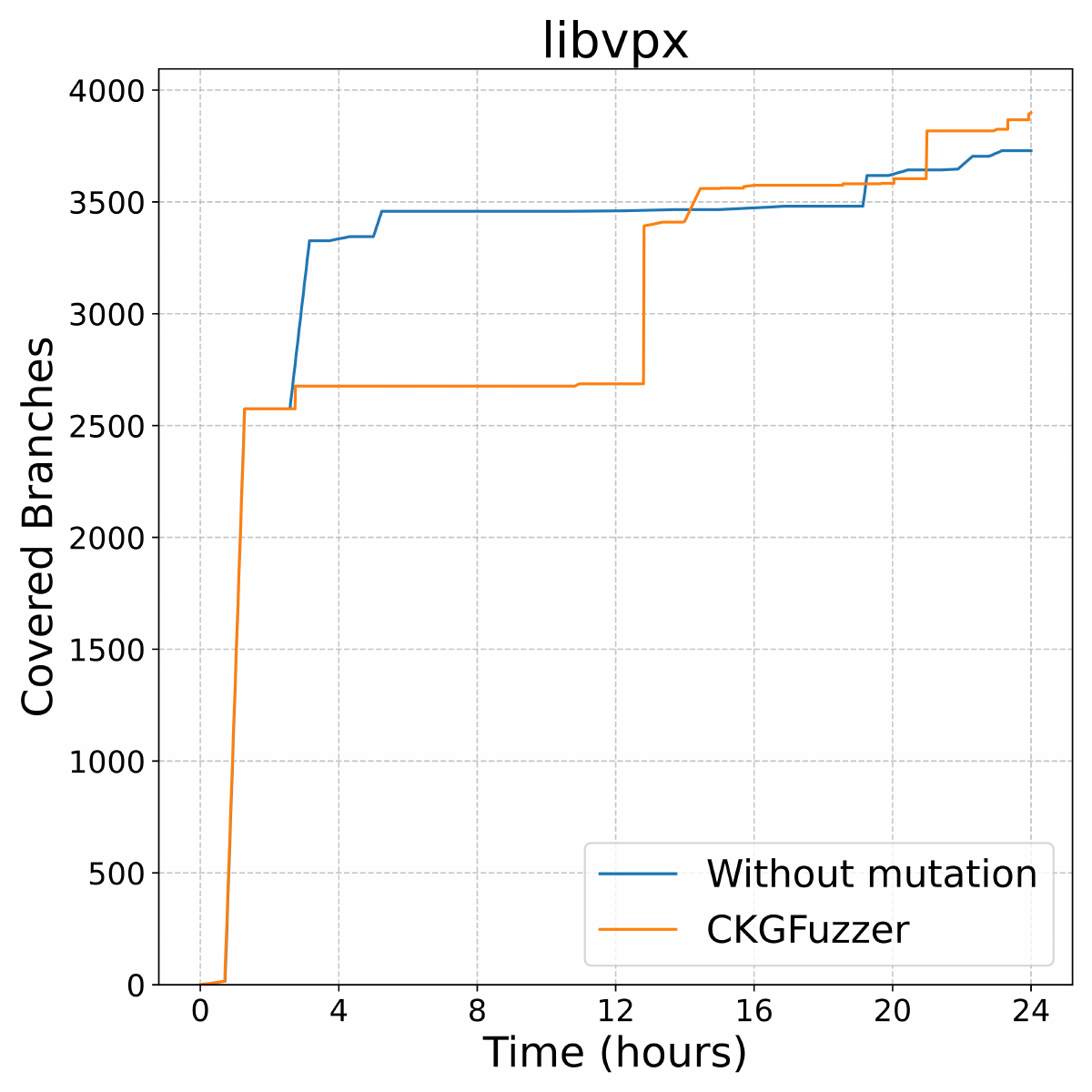}
            \end{subfigure}
            \hfill
            \begin{subfigure}[b]{0.245\textwidth}
                \centering
                \includegraphics[width=\textwidth]{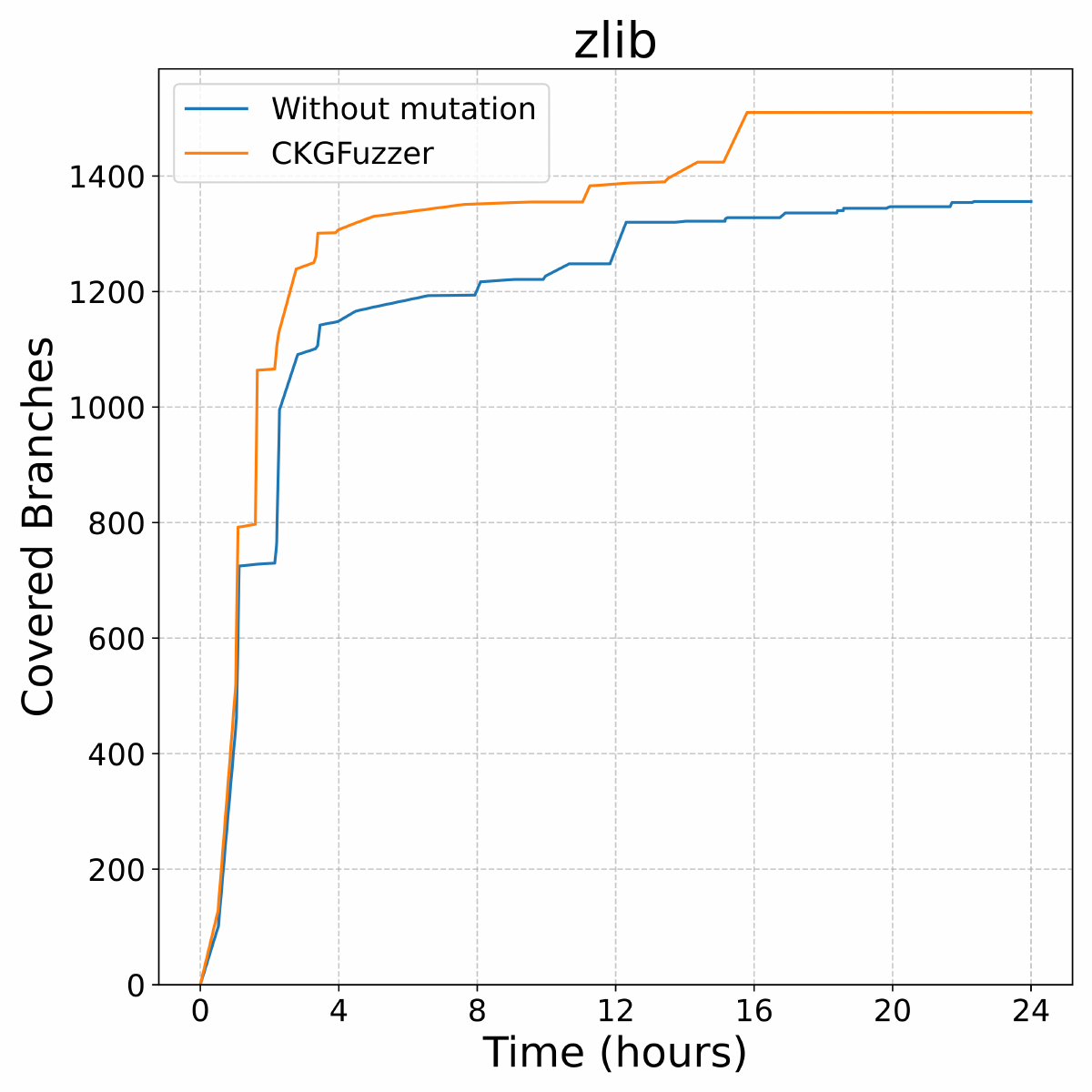}
            \end{subfigure}
        \end{minipage}
    } % End of second row
    \caption{Branch coverage of \approach with and without coverage-guided mutation in different libraries}
    \label{fig:mutation}
\end{figure*}

Additionally, we evaluated the code coverage performance of the two variants. The results in \autoref{tab:comparison} show that both the \texttt{without repair} and \texttt{LLM-only repair} achieved lower code coverage compared to \approach. This is due to many fuzz drivers containing unique API combinations that failed to compile and, thus, could not participate in the fuzzing process, limiting the potential to explore new execution paths in the libraries through diverse API combinations.

\subsubsection{Coverage-guided mutation}
\approach utilizes coverage-guided mutation to discover potential new API combinations. To assess its effectiveness, we conducted a comparative experiment against a variant that does not apply API combination mutation. We kept the fuzzing time fixed at 24 hours for both approaches to compare their coverage trends. \autoref{fig:mutation} shows the branch coverage results of \approach with and without API combination mutation across different libraries.

Our coverage-guided mutation demonstrates a performance advantage across all eight libraries, although the benefit is less pronounced in \texttt{cjson} and \texttt{libvpx}. This is because the number of APIs in these two libraries is relatively small and functionally similar, meaning that nearly all possible API combinations were explored during the initial API combination generation phase. In the remaining six libraries, coverage-guided mutation significantly improved the code coverage of \approach. The experimental results suggest that coverage-guided mutation is a more effective approach in most cases.

\subsection{RQ3. Crash Analysis Result}

\begin{table}[]
\caption{Summary of crash analysis results in \approach}
\begin{tabularx}{\linewidth}{l*{4}{>{\centering\arraybackslash}X}}
\hline
\textbf{Library} & \textbf{UC} & \textbf{MC} & \textbf{RB} & \textbf{CB} \\ \hline
c-ares           & 28          & 23          & 5           & 2(2)        \\
cjson            & 14          & 12          & 2           & 0           \\
curl             & 28          & 24          & 4          & 0           \\
lcms             & 42         & 36          & 6         & 4           \\
libpcap          & 22          & 18          & 4           & 0           \\
libtiff          & 40          & 32          & 8          & 5           \\
libvpx           & 8           & 8           & 0           & 0           \\
zlib             & 17          & 15          & 2           & 0           \\ \hline
\textbf{Total}   & 199         & 168         & 31          & 11(2)       \\ \hline
\end{tabularx}

\vspace{1em}
\raggedright
\footnotesize{\textbf{UC}: Unique Crashes, \textbf{MC}: Misuse Crashes, \textbf{RB}: Reported Bugs, \textbf{CB}: Confirmed Bugs (Fixed)}
\label{tab:bug}
\end{table}

\approach incorporates a crash analysis module to automate the detection of crash causes during fuzzing. As shown in \autoref{tab:bug}, we collected the results from this crash analysis module. 
\subsubsection{Overall result}
\approach identified 199 unique crashes during fuzzing, of which 168 were determined by the crash analysis module to be caused by the misuse of library APIs. To minimize false positives, the crash analysis module treats cases that are ambiguous or difficult to classify as misuse crashes. For example, if a crash is caused by passing a non-existent or error pointer to a library API, and the API does not employ defensive programming to check the validity of the input pointer, we still categorize this as a misuse crash. To verify the accuracy of the crash analysis module in detecting misuse crashes, we randomly sampled 10\% of the misuse crash cases from each library for manual inspection. After reviewing these cases, we confirmed that all of them were indeed caused by the misuse of the API by the fuzz driver.

We also manually reviewed the crashes classified as reported bugs by the crash analysis module. In total, 11 real-world bugs were identified across three libraries, of which nine had not been previously reported. For the cases where the crash analysis module incorrectly classified a crash as a reported bug,  the errors mainly stemmed from complex API call relationships within the libraries. The LLM had difficulty tracing deep nested API calls when parsing the fuzz driver source code, which prevented it from understanding the input constraints and functionality of the nested APIs, leading to incorrect classifications.

Our experimental results demonstrate that the crash analysis module effectively identifies the sources of crashes during fuzzing, reducing the manual review workload by 84.4\%. This significantly improves the efficiency of bug triaging for developers.

\subsubsection{Case study}
\lstset{
  basicstyle=\ttfamily\small,
  keywordstyle=\color{blue!50!black},
  commentstyle=\color{gray},
  stringstyle=\color{green!50!black},
  numbers=left,
  numberstyle=\tiny,
  numbersep=5pt,
  frame=none,
  breaklines=true,
  breakatwhitespace=true,
  showstringspaces=false,
}

\lstdefinelanguage{C}{
  keywords={void, if, else, const, char, tmsize_t, return},
  sensitive=true,
  comment=[l]{//},
  morecomment=[s]{/*}{*/},
  morestring=[b]",
  morestring=[b]'
}

\begin{figure}
\begin{subfigure}{0.5\textwidth}
\includegraphics[width=\textwidth]{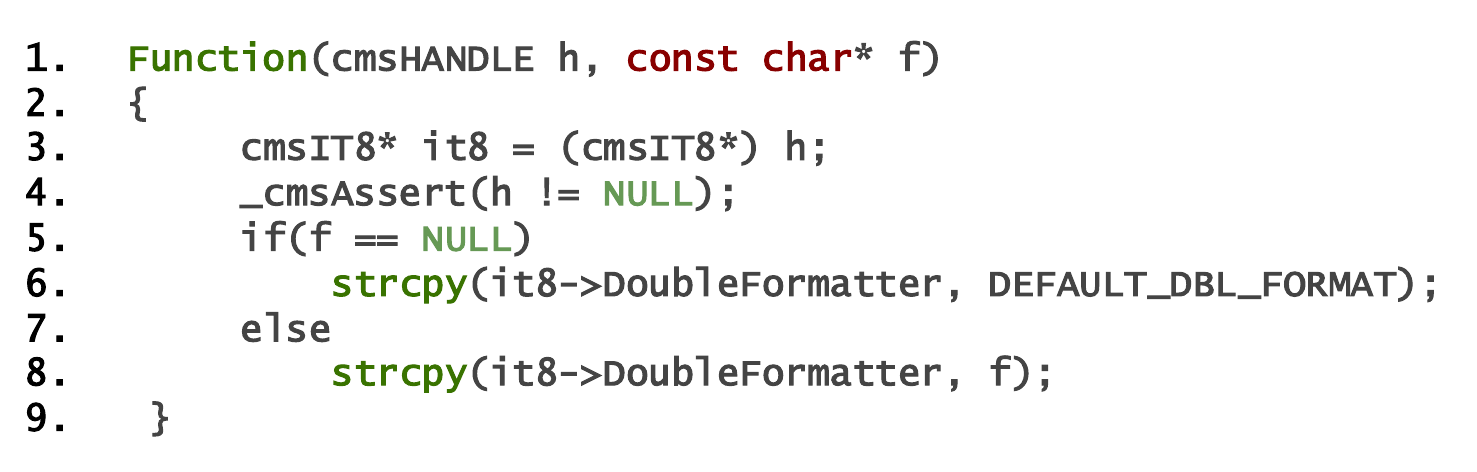}
\caption{A buffer overflow bug case in the \texttt{lcms}}
\label{fig:bug-1}
\end{subfigure}\\
\begin{subfigure}{0.5\textwidth}
\includegraphics[width=\textwidth]{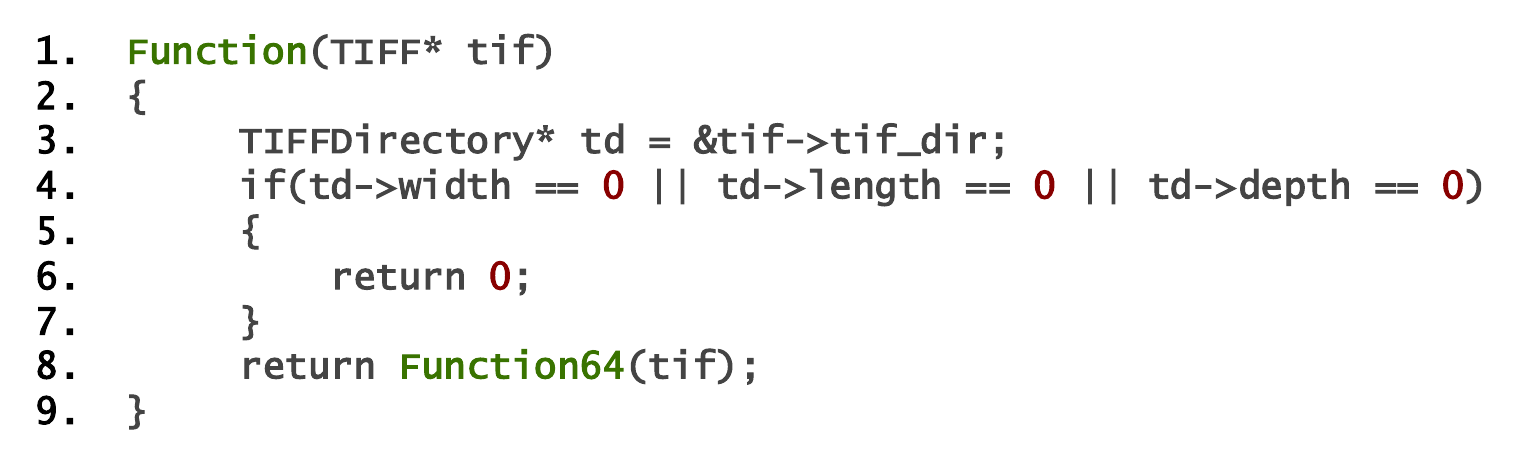}
\caption{An out of memory bug case in the \texttt{libtiff}}
\label{fig:bug-2}
\end{subfigure}
\caption{Bugs found by \approach}
\label{fig:bug}
\end{figure}

\autoref{fig:bug-1} shows a buffer overflow bug case discovered by \approach. In this code, the unsafe \texttt{strcpy} function is used without checking the length of the input string \texttt{f}. Since the data passed by the fuzz driver is randomly generated, the length of \texttt{f} is not controlled, and it may exceed the size of the \texttt{it8->DoubleFormatter} buffer. This leads to a buffer overflow, which is a severe security vulnerability that could potentially be exploited by attackers to overwrite memory and execute arbitrary code.

\autoref{fig:bug-2} illustrates another bug case discovered by \approach, which is an out-of-memory (OOM) issue. In this case, \texttt{*tif} refers to a TIFF file controlled by \approach. The library function does not properly validate the fields such as width and length within the file. As a result, it returns an abnormal tile value. Other library APIs attempt to allocate memory based on the calculated tile size, which causes excessive memory allocation, ultimately leading to an OOM crash. Our case study demonstrates the effectiveness of \approach in detecting real-world bugs. We will publicly disclose the bug cases we have discovered on our website.
\section{Threats to validity}
\subsection{External}
The evaluation of \approach was conducted using a set of eight widely-used open-source libraries. While these libraries cover a range of API usage patterns and functionalities, the generalizability of our approach to other libraries or domains remains uncertain. It is possible that \approach's performance may vary when applied to libraries with different structures, API designs, or use cases. Another potential threat is bias introduced by the training data of the LLM used in \approach. The LLM may favor over-represented patterns in its training set, which could affect the generation of API combinations or the crash analysis process. To mitigate this risk, we employed diverse prompts and knowledge sources to guide the LLM and reduce the impact of any biases.

\subsection{Internal}
A key internal threat to validity is the potential for LLM hallucinations during code generation. Incorrect or irrelevant code snippets may be generated, requiring multiple rounds of refinement or manual intervention to correct. This could affect the quality and functionality of the generated fuzz drivers. Additionally, the performance of \approach depends heavily on the accuracy and completeness of the knowledge graph and API information extracted from the codebase. Incomplete or inaccurate knowledge may result in suboptimal API combinations or ineffective crash analysis. 
\section{Discussion}
\subsection{Applicability to Different Programming Languages}
\approach was primarily evaluated on C-based libraries, which are widely used and have well-documented APIs. However, the applicability of \approach to other programming languages, such as Rust, Java, or Python, remains an open question. Since each language has different API designs, \approach may require adjustments to effectively handle these differences. Future work could explore the adaptation of \approach to these languages, potentially broadening its scope and utility.

\subsection{Quality of the Code Knowledge Graph}
The code knowledge graph plays a central role in \approach by guiding API combination generation and updating API combinations during coverage-guided mutation. The completeness and accuracy of the code knowledge graph are essential for generating effective fuzz drivers. In our experiments, libraries with more comprehensive code knowledge graph showed better coverage and higher fuzzing effectiveness. Improving the automation, construction, and validation of the code knowledge graph could further enhance the performance of \approach by ensuring that it accurately reflects the library's API structure and usage patterns.
\section{Conclusion}
In this paper, we presented CKGFuzzer, a novel LLM-driven fuzz testing framework enhanced by a code knowledge graph. CKGFuzzer automates fuzz driver generation and introduces several key features to improve fuzz testing effectiveness. These include a coverage-guided mutation strategy, which iteratively refines API combinations to explore new execution paths, a dynamic program repair mechanism that automatically resolves compilation errors in generated fuzz drivers, and a crash analysis module that helps identify the root causes of runtime failures. Our experiments on eight open-source projects demonstrate that CKGFuzzer achieves better code coverage and successfully detects 11 real-world bugs. The coverage-guided mutation enables broader exploration of code paths, while the dynamic program repair and crash analysis modules significantly reduce manual intervention, making the fuzzing process more efficient and reliable. These results highlight the potential of leveraging LLMs and code knowledge graphs to advance automated fuzz testing.

%-------------------------------------------------------------------------------
% \section*{Acknowledgments}
% %-------------------------------------------------------------------------------

\newpage
\bibliographystyle{IEEEtranS}
\bibliography{ref}

\end{document}